\input harvmac
\input epsf
\input amssym
%\draftmode
%
%
\noblackbox
%%%%%%%%%%%%%%%%%%%%%%%%%%%%%%%%%%%%%%%%%%%%%%
%%%%%%%%%%%%%%%%%%%%%%%%%%%
% some stuff needed for figures:
%%%%%%%%%%%%%%%%%%%%%%%%%%%%%%%%%%%%%%%%%%%%%%
%%%%%%%%%%%%%%%%%%%%%%%%%%%
\newcount\figno
\figno=0
\def\fig#1#2#3{
\par\begingroup\parindent=0pt\leftskip=1cm\rightskip=1cm\parindent=0pt
\baselineskip=11pt
\global\advance\figno by 1
\midinsert
\epsfxsize=#3
\centerline{\epsfbox{#2}}
\vskip -21pt
{\bf Fig.\ \the\figno: } #1\par
\endinsert\endgroup\par
}
\def\figlabel#1{\xdef#1{\the\figno}}
\def\encadremath#1{\vbox{\hrule\hbox{\vrule\kern8pt\vbox{\kern8pt
\hbox{$\displaystyle #1$}\kern8pt}
\kern8pt\vrule}\hrule}}
%%%%%%%%%%%%%%%%%%%%%%%%%%%%%%%%%%%%%%%%%%%%%%
%%%%%%%%%%%%%%%%%%%%%%%%%%%
% definitions
%%%%%%%%%%%%%%%%%%%%%%%%%%%%%%%%%%%%%%%%%%%%%%
%%%%%%%%%%%%%%%%%%%%%%%%%%%

\def\frac#1#2{{#1 \over #2}}

\def\p{\partial}
\def\semi{\subset\kern-1em\times\;}
\def\bar#1{\overline{#1}}
\def\sqr#1#2{{\vcenter{\vbox{\hrule height.#2pt
\hbox{\vrule width.#2pt height#1pt \kern#1pt \vrule width.#2pt}
\hrule height.#2pt}}}}

\def\p{\partial}

\def\th{\theta}

\def\ad{\bar a}

\def\N{N_{KK}}

\def\rh{\hat{r}}
\def\th{\hat{t}}

\def\rt{{\tilde{r}}}

\def\p{\partial}

%%%%%%%%%%%%%%%%%%%%%%%%%%%%%%%%%%%%%%%%%%%%%%
%%%%%%%%%%%%%%%%%%%%%%%%%%%
% My Definitions
%%%%%%%%%%%%%%%%%%%%%%%%%%%%%%%%%%%%%%%%%%%%%%
%%%%%%%%%%%%%%%%%%%%%%%%%%%
\def\s{\tilde{\cal S}}
\def\nb{\tilde{\alpha}}
\def\Ft{\tilde{F}}
\def\Ih{{\hat I}}
\def\th{{\hat t}}
\def\ih{{\hat i}}

\def\rh{{\hat r}}
\def\ra{\Rightarrow}
\def\half{{1\over 2}}
\def\D{{\cal D}}
\def\N{{\cal N}}

\def\ra{{\rightarrow}}
\def\lt{\left}
\def\rt{\right}
\def\p{\partial}

%%%%%%%%%%%%%%%%%%%%%%%%%%%%%%%%%%%%%%%%%%%%%%
%%%%%%%%%%%%%%%%%%%%%%%%%%%
% References
%%%%%%%%%%%%%%%%%%%%%%%%%%%%%%%%%%%%%%%%%%%%%%
%%%%%%%%%%%%%%%%%%%%%%%%%%%
%\HanakiPJ
\lref\HanakiPJ{
  K.~Hanaki, K.~Ohashi and Y.~Tachikawa,
  %``Supersymmetric Completion of an R^2 Term in Five-Dimensional
  %Supergravity,''
  Prog.\ Theor.\ Phys.\  {\bf 117}, 533 (2007)
  [arXiv:hep-th/0611329].
  %%CITATION = PTPKA,117,533;%%
}

%\LarsenXM
\lref\LarsenXM{
  F.~Larsen,
  %``The attractor mechanism in five dimensions,''
  Lect.\ Notes Phys.\  {\bf 755}, 249 (2008)
  [arXiv:hep-th/0608191].
  %%CITATION = LNPHA,755,249;%%
}

%\CastroSD
\lref\CastroSD{
  A.~Castro, J.~L.~Davis, P.~Kraus and F.~Larsen,
  %``5D Attractors with Higher Derivatives,''
  JHEP {\bf 0704}, 091 (2007)
  [arXiv:hep-th/0702072].
  %%CITATION = JHEPA,0704,091;%%
}

%\CastroHC
\lref\cdkl{
  A.~Castro, J.~L.~Davis, P.~Kraus and F.~Larsen,
  %``5D Black Holes and Strings with Higher Derivatives,''
  JHEP {\bf 0706}, 007 (2007)
  [arXiv:hep-th/0703087].
  %%CITATION = JHEPA,0706,007;%%
}
%\CastroCI
\lref\CastroCI{
  A.~Castro, J.~L.~Davis, P.~Kraus and F.~Larsen,
  %``Precision entropy of spinning black holes,''
  JHEP {\bf 0709}, 003 (2007)
  [arXiv:0705.1847 [hep-th]].
  %%CITATION = JHEPA,0709,003;%%
}

%\CastroNE
\lref\CastroNE{
  A.~Castro, J.~L.~Davis, P.~Kraus and F.~Larsen,
  %``String Theory Effects on Five-Dimensional Black Hole Physics,''
  Int.\ J.\ Mod.\ Phys.\  A {\bf 23}, 613 (2008)
  [arXiv:0801.1863 [hep-th]].
  %%CITATION = IMPAE,A23,613;%%
}

%\CremoniniTW
\lref\CremoniniTW{
  S.~Cremonini, K.~Hanaki, J.~T.~Liu and P.~Szepietowski,
  %``Black holes in five-dimensional gauged supergravity with higher
  %derivatives,''
  arXiv:0812.3572 [hep-th].
  %%CITATION = ARXIV:0812.3572;%%
}

%\CvitanEN
\lref\cpps{
  M.~Cvitan, P.~D.~Prester, S.~Pallua and I.~Smolic,
  %``Extremal black holes in D=5: SUSY vs. Gauss-Bonnet corrections,''
  JHEP {\bf 0711}, 043 (2007)
  [arXiv:0706.1167 [hep-th]].
  %%CITATION = JHEPA,0711,043;%%
}

%\SenWA
\lref\SenWA{
  A.~Sen,
  %``Black Hole Entropy Function and the Attractor Mechanism in Higher
  %Derivative Gravity,''
  JHEP {\bf 0509}, 038 (2005)
  [arXiv:hep-th/0506177].
  %%CITATION = JHEPA,0509,038;%%
}

%\LopesCardosoKY
\lref\card{
  G.~Lopes Cardoso, A.~Ceresole, G.~Dall'Agata, J.~M.~Oberreuter and J.~Perz,
  %``First-order flow equations for extremal black holes in very special
  %geometry,''
  JHEP {\bf 0710}, 063 (2007)
  [arXiv:0706.3373 [hep-th]].
  %%CITATION = JHEPA,0710,063;%%
}

%\PerzKH
\lref\PerzKH{
  J.~Perz, P.~Smyth, T.~Van Riet and B.~Vercnocke,
  %``First-order flow equations for extremal and non-extremal black holes,''
  arXiv:0810.1528 [hep-th].
  %%CITATION = ARXIV:0810.1528;%%
}

%\HyakutakeAQ
\lref\HyakutakeAQ{
  Y.~Hyakutake and S.~Ogushi,
  %``Higher derivative corrections to eleven dimensional supergravity via  local
  %supersymmetry,''
  JHEP {\bf 0602}, 068 (2006)
  [arXiv:hep-th/0601092].
  %%CITATION = JHEPA,0602,068;%%
}

%\CeresoleWX
\lref\CeresoleWX{
  A.~Ceresole and G.~Dall'Agata,
  %``Flow Equations for Non-BPS Extremal Black Holes,''
  JHEP {\bf 0703}, 110 (2007)
  [arXiv:hep-th/0702088].
  %%CITATION = JHEPA,0703,110;%%
}

%\AndrianopoliGT
\lref\AndrianopoliGT{
  L.~Andrianopoli, R.~D'Auria, E.~Orazi and M.~Trigiante,
  %``First Order Description of Black Holes in Moduli Space,''
  JHEP {\bf 0711}, 032 (2007)
  [arXiv:0706.0712 [hep-th]].
  %%CITATION = JHEPA,0711,032;%%
}

%\LiAR
\lref\LiAR{
  W.~Li,
  %``Non-Supersymmetric Attractors in Symmetric Coset Spaces,''
  arXiv:0801.2536 [hep-th].
  %%CITATION = ARXIV:0801.2536;%%
}

%\FerraraDD
\lref\FerraraDD{
  S.~Ferrara and R.~Kallosh,
  %``Supersymmetry and Attractors,''
  Phys.\ Rev.\  D {\bf 54}, 1514 (1996)
  [arXiv:hep-th/9602136].
  %%CITATION = PHRVA,D54,1514;%%
}

%\StromingerKF
\lref\StromingerKF{
  A.~Strominger,
  %``Macroscopic Entropy of $N=2$ Extremal Black Holes,''
  Phys.\ Lett.\  B {\bf 383}, 39 (1996)
  [arXiv:hep-th/9602111].
  %%CITATION = PHLTA,B383,39;%%
}

%\FerraraIH
\lref\FerraraIH{
  S.~Ferrara, R.~Kallosh and A.~Strominger,
  %``N=2 extremal black holes,''
  Phys.\ Rev.\  D {\bf 52}, 5412 (1995)
  [arXiv:hep-th/9508072].
  %%CITATION = PHRVA,D52,5412;%%
}

%\DabholkarTB
\lref\DabholkarTB{
  A.~Dabholkar, A.~Sen and S.~P.~Trivedi,
  %``Black hole microstates and attractor without supersymmetry,''
  JHEP {\bf 0701}, 096 (2007)
  [arXiv:hep-th/0611143].
  %%CITATION = JHEPA,0701,096;%%
}

%\GoldsteinFQ
\lref\GoldsteinFQ{
  K.~Goldstein and S.~Katmadas,
  %``Almost BPS black holes,''
  arXiv:0812.4183 [hep-th].
  %%CITATION = ARXIV:0812.4183;%%
}

%%%%%%%%%%%%%%%%%%%%%%%%
%End Refs
%%%%%%%%%%%%%%%%%%%%%%%%

%%%%%%%%%%%%%%%%%%%%%%%%%%%%%%%%%%
%%%%%%%%%%%%%%%%%%%%%%%%%%%%%%%%%
%Title page, Abstract, and Date
%%%%%%%%%%%%%%%%%%%%%%%%%%%%%%%%%%
%%%%%%%%%%%%%%%%%%%%%%%%%%%%%%%%%
\Title{\vbox{\baselineskip12pt
%\hbox{hep-th/0508218}
%\hbox{UCLA-05-TEP-XX} \hbox{MCTP-XX-XX}
}} {\vbox{\centerline {A Note on Non-BPS Black Holes in 5D}}}
%\medskip\vbox{\centerline {in Theories with Higher
%Derivatives}}} }
\centerline{
Akhil Shah$^\dagger$\foot{akhil137@ucla.edu}}
\bigskip
\centerline{${}^\dagger$\it{Department of Physics and
Astronomy, UCLA,}}\centerline{\it{ Los Angeles, CA 90095-1547,
USA.}}

\baselineskip15pt

\vskip .3in

\centerline{\bf Abstract}
We study the possibility of constructing non-BPS charged black holes in 5D Poincare supergravity by partially violating superconformal Killing spinor equations.  However, solutions to these modified first order  equations are inconsistent with the second order equations of motion beyond the near horizon region.  Instead we find, for special prepotentials, that the consistent asymptotically flat extension of the non-BPS near horizon solution is generated by a symmetry transformation that leaves the two-derivative action invariant.
%%%
\Date{February, 2009}
%%%%%%%%%%%%%%%%%%%%%%%%%%%%%%%%%%%%%%%%%%%%%%
%%%%%%%%%%%%%%%%%%%%%%%%%%%
% Main text begins here
%%%%%%%%%%%%%%%%%%%%%%%%%%%%%%%%%%%%%%%%%%%%%%
%%%%%%%%%%%%%%%%%%%%%%%%%%%
\baselineskip14pt

\newsec{Introduction}
Black hole solutions to supergravity theories coupled to scalar and vector fields have played an essential role in various aspects of string theory, allowing a greater understanding of black hole entropy.  Using the attractor mechanism, \DabholkarTB\ argued for the consistency between microscopic and macroscopic entropy of extremal black holes, both BPS and non-BPS.  The attractor mechanism, initially discovered for supersymmetric (BPS) black holes \refs{\FerraraDD,\StromingerKF,\FerraraIH}, and later found to hold more generally for extremal BPS and non-BPS black holes \SenWA, determines the near-horizon values of scalar fields and black hole entropy solely in terms of the electric and magnetic charges but independently of the asymptotic values of the scalar moduli.  

Although both BPS and non-BPS extremal black holes display the attractor mechanism, explicit BPS solutions, beyond just the near horizon region, are much easier to find as they are determined by solving first-order equations associated with the vanishing of fermionic variations of the supergravity theory\foot{In the presence of higher-derivative corrections, one may need to additionally solve a second order equation of motion \cdkl.}.  Due to the relative difficulty of solving second order equations of motion, any method to obtain explicit non-BPS black holes would be a welcome substitute.  

In four-dimensional $\N=2$ supergravity, first order equations of motions and the conditions for existence of first-order equations were given in \CeresoleWX, for a special class of both BPS and non-BPS stationary extremal black holes.  Subsequently, first order equations were also derived for $d=4$, $\N>2$ theories in \AndrianopoliGT.  In \PerzKH, first-order equations for these extremal\foot{The method in \PerzKH\ also considers non-extremal solutions.} black holes are obtained by rewriting the effective one-dimensional action as a sum and difference of squares, which is possible when the scalar manifold is a symmetric space after timelike dimensional reduction.  Demanding the vanishing of each individual term in the sum of squares effective action, results in first order equations of motion describing the radial evolution of scalars in terms of the metric on moduli space and the Abelian charges of the black hole.  For an alternative approach to certain non-BPS stationary solutions (including multi-center solutions) in four dimensions, see \LiAR\ and references therein.  In five dimensions, \GoldsteinFQ\ give an algorithm to construct stationary non-BPS black holes with a Gibbons-Hawking base space, which are locally BPS, but contain a global obstruction to being supersymmetric.  This approach, however, seems unable to describe asymptotically flat five dimensional non-BPS black holes.  

In this note we will consider an alternative approach to finding non-BPS black holes in 5D using the superconformal formalism.  The superconformal approach to 5D supergravity has allowed the construction of an action with a certain four-derivative term and its supersymmetric completion \HanakiPJ.  Various supersymmetric solutions to this higher-derivative theory have been found and  implications for black hole entropy and the AdS/CFT correspondence have been analyzed \refs{\CastroSD, \cdkl, \CastroCI, \CremoniniTW}.  For a review of this approach see \CastroNE.  However, non-supersymmetric, i.e. non-BPS, solutions to these higher derivative corrections have remain elusive thus far.  The advantage of the superconformal approach is that even in the presence of higher derivative corrections, the Killing spinor equations remain first-order (in contrast to the Noether procedure for constructing higher derivative corrections, see e.g. \HyakutakeAQ) and thus BPS solutions are tractable. 

In \cpps\ a solution was found to the near-horizon equations of motion (i.e. using Sen's entropy function formalism \SenWA) and it was noticed that this solution violated the killing spinor equations, i.e. would be a non-BPS solution.  It is suggestive then to consider whether violating a subset of the Killing spinor equations of the superconformal formalism while satisfying others, allows for consistent non-BPS solutions of the Poincare supergravity.  Instead, we find that the non-BPS configuration of \cpps, when derived by selectively violating some of the Killing spinor equations, is not a consistent asymptotically flat solution to the two-derivative theory, although it is consistent with the equations of motion in the near-horizon region. In the case of a special prepotential (e.g. by reducing M-theory on $K3 \times T^2$), there does exist a consistent asymptotically flat non-BPS solution which reduces to the configuration of \cpps\ in the near-horizon limit.  However this non-BPS solutions is generated by a transformation that leaves the two-derivative action invariant, not by partially violating the Killing spinor conditions. 
 
 \newsec{Near Horizon Solutions}
 We follow the conventions in \cdkl.  The superconformal theory contains a Weyl, $n_V$ vector, and $2r$ hyper multiplets.  The fields in the Weyl multiplet are the vielbein $e^a_\mu$, a two-form auxillary field $v_{ab}$, and an auxillary scalar $D$.  The fields in the vector multiplet are the one-form gauge field $A^I$ and a scalar $M^I$, where $I=1,\ldots,n_V$.  We will not encounter the hypers, as they are only used to gauge fix the dilation symmetry, after which the two-derivative off-shell action becomes
 \eqn\ba{\eqalign{
{\cal L}_0&=-{1\over2}D-{3\over4}R+v^2+{\cal
N}\left({1\over2}D-{1\over4}R+3v^2\right)+2{\cal
N}_Iv^{ab}F^I_{ab}\cr
&~~~~~~~+{\cal
N}_{IJ}\left({1\over4}F^I_{ab}F^{Jab}-{1\over2}\D_aM^I\D^aM^J\right)+{1\over24}
c_{IJK}A^I_{a}F^{J}_{bc}F^{K}_{de}\epsilon^{abcde}.
}}
The geometric data $c_{IJK}$ are the intersection numbers of the Calabi-Yau three-fold on which 11-D supergravity is reduced and define the functions on the scalar manifold \LarsenXM\
\eqn\zxa{
\N={1\over 6}c_{IJK}M^I M^J M^K, \quad \N_I={1\over 2}c_{IJK}M^J M^K, \quad \N_{IJ}=c_{IJK}M^K.
}
The advantage of the superconformal formalism is that the supersymmetry variations of the theory are independent of the action and thus BPS solutions are tractable, even in the presence of higher derivative corrections to \ba.   We consider the following {\it modified} hyperino variation
\eqn\aa{\delta\zeta^\alpha = (\s-1)(\gamma\cdot v)\epsilon^i{\cal A}^\alpha_i,} 
whose vanishing leads to the following conformal-SUSY spinor (parameter of conformal supersymmetry)  
\eqn\ab{\eta^i={\s\over 3}(\gamma\cdot v)\epsilon^i.}  
It will be useful to define another parameter 
\eqn\aba{\nb={2\s \over 3-2\s},}
to distinguish the BPS solutions of \cdkl, which take the values $\s=1$ or equivalently $\nb=+2$, from the non-BPS near-horizon geometry of \cpps, which have $\s=3$ or equivalently $\nb=-2$.  In the rest of the note we will determine which values of $\s$ are allowed in the hopes of finding non-BPS asymptotically flat black holes without directly solving second-order equations of motion.

The remaining supersymmetry variations are
\eqn\ac{\eqalign{
\delta\psi_\mu&=\left({\cal D}_\mu+{1\over2}v^{ab}\gamma_{\mu ab}-{\s\over3}\gamma_\mu\gamma\cdot
v\right)\epsilon~, \cr
\delta\Omega^{I}&=\left(-{1\over4}\gamma\cdot
F^I-{1\over2}\gamma^a\partial_aM^I-{\s\over3}M^I\gamma\cdot v
\right)\epsilon~,\cr
\delta \chi &=\left(D-2\gamma^c\gamma^{ab}{\cal
D}_av_{bc}-2\gamma^a\epsilon_{abcde}v^{bc}v^{de}+
{4\s\over3}(\gamma\cdot v)^2\right)\epsilon~,
}}
where the gaugino index takes $I=0,\ldots,n_V$.  Before seeking asymptotically flat black hole solutions to \ba\ we consider the near horizon region with the following AdS$_2 \times$ S$^3$ ansatz \cpps\
\eqn\ca{\eqalign{
ds^2=v_1\left(x^2 dt^2 - {dx^2\over x^2}\right) &+ v_2 d\Omega^2_3 ~ \Rightarrow R={2(v_2-3v_1)\over v_1 v_2} \cr
~~~~F^I_{tr}(x) = -e^I, \quad & v_{tr}(x) = V \cr
~~~~M^I(x) = M^I, \quad & D(x) = D,
}}
where $R$ is the Ricci scalar, and when substituted in the action \ba\ leads to the entropy function
\eqn\cb{\eqalign{
f_0 = {1\over 4}\sqrt{v_2}&[(\N+3)(3v_1-v_2) - 4V^2(3\N+1){v_2\over v_1} \cr
&+ 8V \N_I e^I{v_2\over v_1}-\N_{IJ}e^I e^J {v_2\over v_1}+ D(\N-1)v_1 v_2].}}
The equations of motion are 
\eqn\cc{\eqalign{
{\partial f_0\over \partial D} &\sim \N-1 = 0\cr
{\partial f_0\over \partial V} &\sim 8\N_I e^I -8V (3\N+1) = 0 \cr
{\partial f_0\over \partial v_1}&\sim 3(\N+3) - {v_2\over v_1^2}\left(-4V^2(3\N+1) +8V\N_I e^I - \N_{IJ} e^I e^J\right)=0\cr
{\partial f_0\over \partial v_2}&\sim 6(v_1-v_2)-{3v_2\over 2v_1}\left(16V^2 -8V\N_I e^I +\N_{IJ} e^I e^J\right) =0,}}
where the last two equations are not independent.  Solving the first gives the special geometry constraint $\N=1$ and the second gives $V=1/4\N_I e^I$.  Using these and adding the second and third equations in \cc, we arrive at
\eqn\cad{16v_1^2 -4v_1 v_2 = 0,}
which implies $v_2=4v_1$ and fixes the relative scales of the near-horizon geometry \ca.  By seeking asymptotically flat solutions to \ac, we will see that \cad\ restricts the allowed valued of $\nb$.  

\newsec{Asymptotic Region}
Let us attempt to find asymptotically flat black holes solutions by  requiring \ac\  to vanish.  A priori, there is no reason to find non-BPS solutions by solving this modified system of first order equations.  There are certain non-BPS black holes in 5D that can be found from first order equations (see for example \PerzKH\ and references therein).  The steps of the calculation are similar to \cdkl.  

\subsec{Gravitino Variation}
Consider the ansatz
\eqn\ad{ds^2=e^{4U_1(x)}dt^2-e^{-2U_2(x)}dx^idx^i,} 
and solve the first equation in \ac\ $\delta\psi_\mu=0$, which gives 
\eqn\ae{U_1={\nb\over 2} U,\quad U_2 = U,\quad v_{ti}={1\over 2}(1+\nb)e^{\nb U}\partial_i U.}
We can find the near horizon geometry by setting $e^{2U}={r^2 \over \ell_s^2}$ in which case we have
\eqn\aea{ds^2=\left({r^2\over \ell_s^2}\right)^{\nb} dt^2 - \left({r^2\over \ell_s^2}\right)^{-1}dr^2 - \ell_s^2 d\Omega^2_3.}
For general $\nb$, the following change of coordinates
\eqn\aeca{r^2=(\nb)^{-2/\nb}~(\ell_s)^{2+2/\nb}~(z)^{-2/\nb},}
gives rise to a AdS$_2 \times$ S$^3$ geometry with arbitrary relative scale
\eqn\aecb{ds^2={\ell_s^2\over \nb^2z^2}(dt^2 - dz^2) - \ell_s^2 d\Omega^2_3.}
To bring \aecb\ into the form used in the entropy function of \cpps, we can again change coordinates, $z^2=\ell_s^4/\rho^2$, followed by rescaling, $\rho = \ell_s x$ and $t=\ell_s \tau$, which gives us the metric 
\eqn\zxc{ds^2={\ell_s^2\over\nb^2}(x^2 d\tau^2-{1\over x^2}dx^2)-\ell_s^2d\Omega_3^2,}  
and comparing to the near-horizon ansatz \ca, we have $v_2=\ell_s^2$.  Recall that in the near-horizon analysis of the previous section, the equations of motion (entropy function) resulted in \cad, fixing the relative scale of AdS$_2$ and S$^3$.  The near-horizon restriction of $v_2=4v_1$ translates into $\nb=\pm 2$ as the only allowed values.  These two special values of $\nb$ are the ones considered in \cpps, and each with an appropriate change of coordinates from \aeca, 
\eqn\aeb{\eqalign{
\nb&=+2~, r^2={\ell_s^3\over 2z}~\cr
\nb&=-2~, r^2={\ell_s z},}}
results in the same extremal AdS$_2 \times S^3$ near-horizon geometry 
\eqn\aec{ds^2={\ell_s^2\over 4z^2}(dt^2 - dz^2) - \ell_s^2 d\Omega^2_3.}
We can also check the near-horizon values of the auxillary two-form by noting the general result from the time component of $\delta\psi_t=0$ (assuming $v^{ij}=0$ for a spherically symmetric solution)
\eqn\aed{v^{\th\ih}={3\over 4}{\omega^{\th\ih}_t\over \s e^{\th}_t},} 
and using the spin-connection for the near horizon geometry
\eqn\aee{\omega^{\th\ih}_t=-r.}
After converting to coordinate frame and using the notation of \cpps, we arrive at the near horizon result 
\eqn\aef{v_{ti}={3\sqrt{v_1}\over 4\s}~={3 \ell_s\over 8\s}.}

\subsec{Auxiliary Fermion variation}
Solving the last variation in \aa, we get the equation for the auxiliary scalar
\eqn\al{D=(1+\nb)e^{2U}\left(\nabla^2 U - 2(1+\nb)(\nabla U)^2\right).}  
To get the near horizon value, note that the covariant derivative of $v_{ab}$ in \ac\ must vanish in the extremal background and thus the only remaining term is $\sim \s(\gamma\cdot v)^2$ which scales with $\s$ just like $v_{ab}$ in \aef.  Given the vanishing covariant derivative, the remaining terms in the last equation of \ac\ imply
\eqn\ala{D=-{4\s\over 3}(\gamma\cdot v)^2=-{16\s\over 3}v_{\th\rh} v_{\th\rh}=-{3\over\s v_1},}
which agrees with both the BPS and non-BPS near horizon values of \cpps.  

Summarizing the results so far, we see that the asymptotically flat ansatz \ad\ and vanishing of the modified variations \ac\ lead to the full radial dependence of the auxiliary two-form \ae\ and auxillary scalar \al, whose near-horizon limits are \aef\ and \ala\ respectively.  By taking into account the restriction on the relative scales of the near-horizon AdS$_2 \times$ S$^3$ geometry \cad, we are only allowed $\nb=2$ (i.e. $\s=1$) or $\nb=-2$ (i.e. $\s=3$), leading to the two choices of near-horizon fields defined in \ca\
\eqn\zxd{\eqalign{
\nb=+2 &\Rightarrow v_2=4v_1, \quad D=-{3\over v_1}, \quad V={3\over 4}\sqrt{v_1} \cr
\nb=-2 &\Rightarrow v_2=4v_1, \quad D=-{1\over v_1}, \quad V={1\over 4}\sqrt{v_1}.
}}
In \cpps\ the near-horizon configurations \zxd\ were found in the restricted case of the STU prepotential (i.e $n_V=3$ and $\N=M^1 M^2 M^3$), whereas here we have not assumed any restriction on the geometric data $c_{IJK}$.

\subsec{Gaugino Variations}
If we try to similarly solve the second equation in \ac\ for all values of the index $I$ of the gaugino variation, we will find an inconsistency with the EOM for the auxillary field $v_{ab}$.  We can show this generally by considering the linear combination
\eqn\af{M_I \delta\Omega^{I}=\gamma\cdot\left(-{1\over 4} M_I F^I -\s v\right)\epsilon.}  If we were now to assume that $\delta\Omega^{I}=0$ for all values of the index $I$ then the above linear combination would identically vanish and force us to have 
\eqn\ag{\delta\Omega^{I}=0 \ra v_{ab}=-{1\over 4\s}M_I F^I_{ab},}
which is only consistent with the EOM for $v_{ab}$ when $\s=1$ ( i.e. when the hyperino variation in \aa\ is zero).  To find non-BPS solutions we thus have to ensure that not all the variations of the gaugino are identically zero.  Let us simply take 
\eqn\ah{\delta\Omega^{0}\neq 0,\quad \delta\Omega^{\Ih}=0,}
where $\Ih=1,\ldots,n_V$.  At the same time, we ensure that the EOM for the auxillary two-form is satisfied, i.e. $v_{ab}=-{1\over 4}M_I F^I_{ab}$.  This will allow us to solve for all the gauge fields.  

Again consider the linear combination of \af, in which the left hand side becomes simply $M_0\delta\Omega^{0}$ due to \ah.  Thus our result for the non-vanishing gaugino variation is 
\eqn\ai{\delta\Omega^{0}=\gamma\cdot{\left(-{1\over 4} M_I F^I -\s v\right)\over M_0}\epsilon.}
Solving the vanishing variations in \ah, we have
\eqn\aj{F^\Ih_{it}=\partial_i(e^{\nb U} M^\Ih),}
which we can use to solve the non-vanishing variation in \ah, resulting in
\eqn\ak{\eqalign{
F^0_{it}&={6(1-\s)\over 3-2\s}e^{\nb U}{\partial_i U \over M_0} + \partial_i(e^{\nb U} M^0) \cr
&=(2-\nb)e^{\nb U}{\partial_i U \over M_0} + \partial_i(e^{\nb U} M^0).}}
As a check on our solutions we note that we recover the results in \cdkl\ (eqn 4.18) in the $\s=1\ra\nb=2$ case and we can also verify that with the above solutions \aj, \ak\ for the gauge fields, the equation of motion for $v_{ab}$ is satisfied.  The following notation for the field strengths will be more useful
\eqn\aka{\Ft^I_{it} =F^I_{it} + \delta^{I,0} (2-\nb)e^{\nb U}{\partial_i U \over M_0},} 
where $F^I_{it}$ is the field strength in the supersymmetric case (i.e. as in \aj\ but without the restriction on the vector multiplet index).  We will comment on how this changes the conserved charges when we consider Maxwell's equation.

Note with $\nb=+2$, all the variations in \ac\ and \aa\ vanish, and thus will correspond to a BPS solution, which can be fully extended to the asymptotically flat region as in \cdkl.  In the next section we see whether the non-BPS solution to the modified variations \ac\  (with $\nb=-2$) is consistent with the two-derivative equations of motion to \ba, beyond the near-horizon limit.  

\newsec{Equations of Motion}
We will compute Maxwell's equation and compare it (or a contraction of it) to the time-time component of Einstein's equation in the case of a prepotential gotten by reducing 11D supergravity on K3 $\times$ T$^2$.  

\subsec{Maxwell's Equation}
The two derivative action is 
\eqn\ba{\eqalign{
{\cal L}_0&=-{1\over2}D-{3\over4}R+v^2+{\cal
N}\left({1\over2}D-{1\over4}R+3v^2\right)+2{\cal
N}_Iv^{ab}F^I_{ab}\cr
&~~~~~~~+{\cal
N}_{IJ}\left({1\over4}F^I_{ab}F^{Jab}-{1\over2}\D_aM^I\D^aM^J\right)+{1\over24}
c_{IJK}A^I_{a}F^{J}_{bc}F^{K}_{de}\epsilon^{abcde}
}}
Now we solve Maxwell's equation.  First we need the electric fields (note below we use $F_{ti}=-F_{it}$) for which we use \aka
\eqn\am{\eqalign{
{\cal E}^i_I &= {\p{\cal L}\over \p F^I_{ti}}\cr
&=4M_I v^{ti}+{\cal N}_{IJ}F^{Jti} \cr
&=g^{tt}g^{ii}({\cal N}_{IJ}-M_I M_J)F^J_{ti} \cr
&= e^{U(4-\nb)}\left(\partial_i(e^{-2U}M_I)+e^{-2U}{{\cal N}_{I0}\over M_0}(2-\nb)\partial_i U\right)
}.} 
If we now specialize to the $K3 \times T^2$ prepotential using the following identities
\eqn\amx{\eqalign{
\N_0 = c_{ij}M^i M^j, \quad \N_i &= 2M^0 c_{ij}M^j, \quad \N_{00}=0, \quad \N_{0i}=2c_{ij}M^j, \quad \N_{ij}=2M^0 c_{ij}\cr
~~~~~~~~~~~~~~~~~~~~~~~~\N_0 M^0 &=1, \quad \N_i M^i = 2,
}}
we can obtain a relation on the moduli by contracting Maxwell's equation with the scalars (also multiply by $e^{2U}$ to get rid of exponential factors)
\eqn\amy{\eqalign{
M^I &\p_i (\sqrt{g}{\cal E}^i_I)e^{2U} = 0\cr
&=M^I \nabla^2 \N_I - (4\nabla\N_0 M^0 + (2+\nb)\nabla\N_\Ih M^\Ih)\nabla U \cr
&~~~~~~~+(4M^0\N_0 + 2\nb M^\Ih \N_\Ih)(\nabla U)^2 \cr
&~~~~~~~~~~~~ -(2\N_0 M^0 + \nb N_\Ih M^\Ih)\nabla^2 U \cr
&=M^I \nabla^2 \N_I - (2-\nb)\nabla U M^0 \nabla\N_0 + 4(1+\nb)(\nabla U)^2-2(1+\nb)\nabla^2 U.
}}
Using the result $M^I\N_I = {\rm const}$, we can rewrite \amy\
\eqn\amz{
0=-\half \N_{IJ}\nabla M^I \nabla M^J -\half(2-\nb)\nabla U M^0 \nabla\N_0 + 2(1+\nb)(\nabla U)^2-(1+\nb)\nabla^2 U,
}
which we will compare below to the time-time component of Einstein's equation.

%
%Noting the density $\sqrt{g}=e^{U(\nb-4)}$ in our metric, Maxwell's equation becomes
%\eqn\an{\eqalign{e^{2U}\partial_i(\sqrt{g}{\cal E}^i_I) &=0=\cr
%{\cal N}_{IJ}\partial_i^2 M^J + & c_{IJK}\partial_i M^J \partial_i M^K \cr
%&~~~+\nabla^2 U\left\{-2{\cal N}_I +(2-\nb){{\cal N}_{I0}\over M_0}\right\}\cr
%&~~~+(\nabla U)^2\left\{4{\cal N}_I - 2(2-\nb){{\cal N}_{I0}\over M_0}\right\}\cr
%&~~~+\nabla U \left\{-4{\cal N}_{IJ} + (2-\nb)\left({c_{IJ0}\over M_0}-{{\cal N}_{I0}{\cal N}_{J0}\over M_0^2}\right)\right\}\partial_i M^J = 0.
%}}
%

\subsec{Einstein's equation}
The two-derivative Einstein equations can be written as
\eqn\xah{
-{1\over4}\lt[\lt( \N +3 \rt) \lt(R_{ab}-\half g_{ab}R\rt) -\lt(\D_a \D_b -g_{ab}\rt) \N\rt]+\tilde{T}_{ab} -\half g_{ab}\tilde{\cal L}=0
}
where $\tilde{\cal L}$ is the Lagrangian \ba\ without the Ricci scalar and Chern-Simons terms
\eqn\xai{
\tilde{\cal L} = -{1\over2}D+v^2+{\cal
N}\left({1\over2}D+3v^2\right)+2{\cal
N}_Iv^{ab}F^I_{ab}+{\cal
N}_{IJ}\left({1\over4}F^I_{ab}F^{Jab}-{1\over2}\D_aM^I\D^aM^J\right)
}
and $\tilde{T}_{ab}$ is (sort of) the matter stress tensor
\eqn\xaj{
\tilde{T}_{ab} \equiv {\p \tilde{\cal L} \over \p g^{ab}}= 2\lt(1+3\N\rt)v_{ac}v_b^{~c} +2 \N_I \lt(v_a^{~c}F^I_{bc} + v_b^{~c}F^I_{ac}\rt)+\half{\cal
N}_{IJ}\left(F^I_{ac}F^J_{bd}g^{cd}-\D_aM^I\D_bM^J\right) 
}
We will use the special geometry constraint $\N=1$ valid at the two derivative level.  Defining the usual Einstein tensor $G_{ab}\equiv R_{ab}-\half g_{ab}R$, the time-time component of \xah\ for our general metric \ad\ is 
\eqn\cb{\eqalign{
-G_{tt}&+{\tilde T}_{tt} - {1\over 2}g_{tt}{\tilde {\cal L}}= 0\cr
&=-[-3(\nabla U)^2 + 3\nabla^2 U]+ (\nb^2 + 2)(\nabla U)^2 + (2-\nb)\N_0\nabla U \nabla M^0 - {1\over 2}\N_{IJ}\nabla M^I \nabla M^J \cr
&~~~~~~~~~~~~~~~~~~- {1\over 2}\lt[(\nb^2 + 2)(\nabla U)^2 + (2-\nb)\N_0\nabla U \nabla M^0\rt]\cr
&=- {1\over 2}\N_{IJ}\nabla M^I \nabla M^J - 3\nabla^2 U + ({1\over 2}\nb^2+4)(\nabla U)^2 +{1\over 2}(2-\nb)\N_0\nabla U\nabla M^0
}}

\subsec{Summary}
In the case of the K3 $\times$ T$^2$ prepotential \amx, we see that Maxwell's equation \cb\ and a component of Einstein's equation \amz\ are inconsistent unless $\nb=+2$, which is the BPS solution found in \cdkl.  In the near horizon region $\nb=-2$ can also be a solution, because as we saw previously the near-horizon geometry still remains AdS; only $v,D$, and the relation between $F^0$ and $M^0$ are changed.  In \cpps\ these non-BPS solution for $v,D$, and $F$ were noticed to satisfy only a subset of the BPS equations \ac\ with the paramter $\nb=-2$ (resp. $\s=3$), and were consistent with the near-horizon equations of motion for the fields as indicated in \aef,\ala.   Beyond the near-horizon region, the field configurations that make the variations in \ac\ vanish for $\nb=-2$ are not consistent with the equations of motion.

Instead we will see that the near-horizon non-BPS solution suggested in \cpps\ are gotten by taking the near horizon limit of a solution which is generated by a field transformation on the auxillary two-form $v$ and field strength $F$ that leave the two-derivative action invariant,  but still satisfy $\nb=+2$ .  These field transformations will only coincide with taking $\nb=-2$ for the near horizon limits of $v$ and $F$.

\newsec{Non-BPS solutions from Symmetries of Action}
In \card\ stationary non-BPS extremal BH solutions (generally rotating) at the two-derivative level are found for special cases of prepotentials.  When considering a static BH in the case of the a K3$\times$ T$^2$ prepotential ($\N=M^0\half c_{ij}M^i M^j$), the only difference between the BPS and non-BPS solution is the rotation of the charge vector; i.e., flipping the sign for the $q_0$ charge.  Let us note that with this prepotential, flipping the sign for the $q_0$ charge is a  symmetry of the two derivative action \ba.  Ignoring the CS term, the two-derivative action for this prepotential can be written as
\eqn\ce{
\eqalign{
{\cal L}_0&={1\over2}({\cal N}-1)D-{1\over4}({\cal N}+1)R- {1\over2}{\cal N}_{IJ}\D_aM^I\D^aM^J\cr
&~~~~~~~+(3{\cal N}+1)v^2+2{\cal
N}_0v^{ab}F^0_{ab}+2{\cal
N}_iv^{ab}F^i_{ab}
+{1\over2}{\cal
N}_{0i}F^0_{ab}F^{iab}+{1\over4}{\cal
N}_{ij}F^i_{ab}F^{jab}. 
}
}
Consider the following transformation
\eqn\cf{
F^0 \ra -F^0, \quad v'= v + {1\over 2}{\cal N}_0 F^0,
}
which results in the following change in \ce\ 
\eqn\cg{
\delta {\cal L}_0\equiv {\cal L}_0(v', -F^0)-{\cal L}_0(v,F^0) = ({\cal N}_0{\cal N}_i - {\cal N}_{i0})F^iF^0 + {({\cal N}-1)\over 4}F^0{\cal N}_0(3F^0{\cal N}_0+12v).
}
Now note that the first term in \cg\ vanishes due to the properties of the assumed prepotential 
\eqn\ch{
{\cal N}_{0i} = 2c_{ij}M^i, \quad {\cal N}_i = 2M^0 c_{ij}M^j, \quad {\cal N}_0 M^0=1. 
}
The second term vanishes identically at the two-derivative level since the EOM for $D$ guarantees ${\cal N}=1$.

Flipping the sign of the charge really means changing the relationship between electric fields and the scalar moduli.  Namely in the conventions of \cdkl, a non-BPS black hole solution for the above prepotential would be 
\eqn\cc{
F^\Ih_{it} = + \partial_i(e^{2U} M^\Ih),\quad F^0_{it}= - \partial_i(e^{2U} M^0).  
}
as opposed to \aka\ found by solving the system of first order Killing spinor equations \ac\ with general $\nb$.  In the special case of the K3 $\times$ T$^2$ prepotential, \cpps\ find a solution to the near-horizon field equations that corresponds to $M^\Ih = +e^\Ih$ and $M^0 = -e^0$, for the two derivative case, which we see is exactly the near-horizon limit of \cc, and since this solution is generated by a symmetry of the action, it is perfectly consistent with the equations of motion. 

Interestingly the near-horizon limits of both the consistent \cc\ and inconsistent \aka\ gauge fields result in the non-BPS  values ($\nb=-2$) for the auxiliary fields \zxd, when the prepotential is restricted to that of K3 $\times$ T$^2$ \amx.  Let us detail the calculation for $I=0$ multiplet gauge-field in the inconsistent solution \aka, noting that in the near-horizon region $M^I={\rm constant}$, and the metric takes the form \aea\
\eqn\zza{\eqalign{
F^0_{\ih\th}&=-e^{U}e^{+2U}F^0_{it} \cr
&=-{r\over \ell_s}\left({4\over M_0}{1\over r}-2{M^0\over r}\right) \cr
&=-{2M^0\over \ell_s} \cr
&=-v_1^{-1/2}M^0,
}}
where we have used that $M_0M^0=1$ from \amx\ and $\ell_s=2v_1^{1/2}$ from \cad.  Note the same near horizon limit results for the consistent gauge field $F^0$ in \cc.  After reverting to curved indices the final result for the near-horizon limits for the both the consistent \cc\ and inconsistent \aka\ gauge fields is 
\eqn\zzb{
F^{\Ih}_{it}=-v_1^{1/2}M^\Ih, \quad F^0_{it}=+v_1^{1/2}M^0.
} 
Using the field equation for the auxiliary two form 
\eqn\zzc{\eqalign{
v_{it}&=-{1\over 4}M_I F^I_{it} \cr
&=-{1\over 4}v_1^{1/2}(M_0M^0-M_\Ih M^\Ih) \cr
&={v_1^{1/2}\over 4},
}}
we recover the $\nb=-2$ non-BPS near-horizon value for the auxillary two-form \zxd.  The result for the non-BPS auxiliary scalar $D$ is related to the non-BPS auxillary two-form $v_{ti}$, by using the equation of motion for the hyper scalar $D=2v^2-3/2{\cal R}$, where ${\cal R}$ is the Ricci scalar of \ad\ (for $\nb=+2$, since the consistent solution \cc\ result from symmetry transformations that do not affect the metric from it's BPS configuration).  Taking the near-horizon limit recovers \zxd\ for $\nb=-2$.  

\newsec{Conclusion}
In this note we considered a method to find non-BPS asymptotically flat black hole solutions to five dimensional Poincare supergravity coupled to multiple Abelian vectors and neutral scalars, based on the idea of violating some of the Killing spinor conditions of a theory with a larger symmetry group, namely superconformal supergravity.  A priori, there is no reason such a method should result in a consistent solution, but we were motivated by the near horizon solutions found in \cpps, which noticed such a partial violation of the superconformal Killing spinor constraints.  However, our attempt to explicitly construct the asymptotically flat solution beyond the near-horizon region, following the approach in \cdkl, found inconsistencies with equations of motion of the two-derivative theory.  

In summary we note that the near-horizon non-BPS solution of \cpps, is not the result of solving the modified system of first order Killing spinor equations \ac\ and then taking the near-horizon limit, but is really the near horizon limit of a non-BPS solution that is generated by a symmetry of the action for a special prepotential.  It is not obvious that the symmetry transformation for the special prepotential holds beyond the two-derivative theory.  For generic prepotentials, a systematic method for finding non-BPS solutions remains an interesting open problem.

%%%%%%%%%%%%%%%%%%%%%%%%%%%%%%%%%
%Acknowledgments
%%%%%%%%%%%%%%%%%%%%%%%%%%%%%%%%%
\bigskip
\noindent {\bf Acknowledgments:} \medskip \noindent 
I thank Josh Davis and Per Kraus for very useful discussions.  A.S. was partially supported by a NSF Graduate Fellowship.
%%%%%%%%%%%%%%%%%%%%%%%%%%%%%%%%%%
%%%%%%%%%%%%%%%%%%%%%%%%%%%%%%%%%

\listrefs
\end